\documentclass[12pt]{article}
\def\be{\begin{equation}}
\def\ee{\end{equation}}
\def\bea{\begin{eqnarray}}
\def\eea{\end{eqnarray}}

\def\a{\alpha}
\def\b{\beta}
\def\g{\gamma}
\def\d{\delta}
\def\e{\epsilon}
\def\f{\phi}
\def\k{\kappa}
\def\l{\lambda}

\def\r{\rho}

\def\th{\theta}
\def\x{\xi}
\begin{document}
\begin{center}
\vspace{.8cm}
\Large
{\bf REDSHIFTS and KILLING VECTORS}\\
\vspace{.5cm}
\large{Alex Harvey}${}^{a)b)}$ \\
\vspace{.15cm}
\normalsize
Visiting Scholar \\
New York University \\
New York, NY 10003 \\
\vspace{.5cm}
\large{Engelbert Schucking}${}^{c)}$  \\
\vspace{.15cm}
\normalsize
Physics Department \\
New York University \\
4 Washington Place \\
New York, NY 10003 \\
\vspace{.5cm}
\large{Eugene J. Surowitz}${}^{d)}$ \\
\vspace{.15cm}
\normalsize
Visiting Scholar \\
New York University \\
New York, NY 10003 \\
\vspace{.8cm}
\end{center}
\begin{abstract}
Courses in introductory special and general relativity have
increasingly become part of the curriculum for upper-level
undergraduate physics majors and master's degree candidates.  One of
the topics rarely discussed is symmetry, particularly in the theory of
general relativity.  The principal tool for its study is the Killing
vector.  We provide an elementary introduction to the concept of a
Killing vector field, its properties, and as an example of its utility
apply these ideas to the rigorous determination of gravitational and
cosmological redshifts.
\end{abstract}
\parskip=0.2in
\parindent=1cm
\eject
\section{Prologue}
\setcounter{equation}{0}
In 1907, Albert Einstein enunciated his \lq\lq equivalence
principle\rq\rq\ and used it to examine the influence of the
gravitational field on the propagation of light${}^1$.  He demonstrated, 
in approximation that as light moved through a difference in 
gravitational potential its frequency would change.  It is with the latter
that the concept of {\it gravitational redshift\/} enters the ken of the
physicist.

In 1916 Einstein completed his journey to a theory of gravitation:
{\it The Foundations of the General Theory of Relativity\/}${}^2$.
Here, redshift is studied with greater precision.  It is
discussed in terms of the {\it metric}.  In Section 22, {\it Behavior
of Rods and Clocks in the Static Gravitational Field}, toward the end
of the paper he states, (in translation)
\lq\lq From this it follows that the spectral lines
of light reaching us 
from the surface of large stars must appear
displaced toward the red end of the spectrum.\rq\rq

The General Theory of Relativity describes gravitation as
the curvature of spacetime.
The Einstein equations are a protocol for the determination,
subject to hypotheses concerning the physical nature of the spacetime being discussed,
of the metric of the space in question.
This metric is the source of {\it all\/} information about the properties of the space.
In particular, of special interest here,
it enables the calculation of the paths of
light rays---the null geodesics---and the description of the world lines
of objects in the spacetime.

These ideas were substantially refined and put into the form in use today by the distinguished mathematician Hermann Weyl.  He introduced the concept of world lines of emitter and observer
and the connection of elements of proper time on each defined by null geodesics propagating from the former to the latter.  This, for the first time, provided
a general definition of frequency shift for light in arbitrary
spacetimes.  It was first stated in the fifth edition of
his book Space-Time-Matter${}^3$.
In Appendix III \lq\lq Redshift and Cosmology\rq\rq\ he explains:

\begin{quote}
\lq\lq The different points
on the world line of a point-like light source are the origin of
(three-dimensional) surfaces of constant phase that form null cones
opening towards the future. From the rhythm of the change of phase on
this line one obtains the perceived change of phase for any observer
by checking how his world line intersects the successive surfaces of
constant phase.  Let $s$ be the proper time of the light source and
$s'$ the proper time of the observer. To each point $s$ on the world
line of the light source corresponds a point $s'$ on the world line of
the observer: $s' = s'(s)$, the intersection of his world line with
the future null cone emanating from $s$.  If the process occurring at
the position of the source is a purely periodic one of infinitely
small period then the change of phase experienced by the observer is
also periodic; but the period is increased in the ratio
$1 + z = ds'/ds$
(obviously measured along both world lines in their respective
proper times).  If the observer carries a light source of the same
physical nature as the observed one then every spectral line of his
light source with frequency $f$ corresponds to the spectral line of the
distant light source with frequency $ f/(1+z) $.  The ones appear to be
displaced with respect to the others.\rq\rq
\end{quote}

In the ray approximation, light source and observer are connected by
null geodesics running on surfaces of constant phase of the future
light cones. (See Fig.~1).
Null geodesics emitted at proper times $s$ and $s + ds$
will intersect the world line of the observer at proper times $s'$ and
$s' + ds'$ respectively. (See Fig.~2). The redshift is simply
\be
\label{weyl}
                  1 + z = \begin{array}{c} ds' \\ \overline {ds} \end{array}\, .
\ee 
In order to find this ratio one must know the
null geodesics, or more precisely, their variation.  In general this
is not simple.  The task is greatly simplified if, under a time
translation, the geodesic is not changed or is subjected only to a
scale transformation.  {\it These are the conditions, respectively,
for the existence of either a Killing vector or a conformal Killing
vector in the spacetime manifold.}  This is the central point of our
investigation.
\section{Introduction}
\setcounter{equation}{0}
The concept of symmetry is central to the solution of many problems in physics.
Introduction of ignorable coordinates in the construction of kernel functions for Lagrangians
entails implicit use of {\it a priori\/} knowledge of the symmetries of the system being studied.
Construction of the Hamiltonian for a problem in quantum mechanics is constrained by the symmetries
thereof.  It is the measure of the power of symmetry 
considerations that solely through their application, the Robertson-Walker metric can be constructed.
In this paper we study redshift and, as we shall show, the symmetries of the spacetime determine the manner in which redshifts occur.

Spacetime is a 4-dimensional Riemannian manifold, that is, it is a surface.  
The symmetries of a surface are numerous: among these there are discrete symmetries such as
inversion in a point or reflections in a plane and there are the infinitesimal continuous coordinate transformations (also called mappings or motions) which {\it leave the
metric unchanged}.  Such mappings are called \lq\lq isometries\rq\rq.  It is the latter which will occupy our attention.

A significant, yet obvious example is furnished by the isometries of the metric
of special relativity.
Its \lq\lq flatness\rq\rq\ enables their simple enumeration: 3 spatial translations,
3 rotations, 3 pure Lorentz transformations, and the translations
along the time axis.  This last is worth noting.  In a 4-dimensional space, time is the same as
any other coordinate.  If a metric is invariant under translations in the $x$-direction this is
no different in principle than saying that a metric is invariant under translations along the
$t$-axis.  The latter implies that the metric is stationary.  Conversely, if a metric is stationary it
possesses a time symmetry.

Other well-known examples are the sphere which
has the symmetry of the well-known three-dimensional
rotation group, ${\bf O}_3$.  In the curved spacetime of general
relativity things are rarely that obvious.  The continuous symmetries
are those of interest here.

The principal tool for investigating the isometries of a metric
is the Killing vector field${}^{4,5}$ which
was introduced by the late 19th century German
mathematician Wilhelm Killing, its distinguished eponym.  It was developed and
exploited in the study of continuous groups.  The set of Killing
vectors for a given metric provides an invariant characterization of
these properties.
No matter the coordinate system in which the metric is cast,
its set of Killing vectors ({\it modulo\/} coordinate
transformations) will be the same.
\section{Killing Vector Fields}
\setcounter{equation}{0}
Consider the infinitesimal change in a metric, $g_{ab}$, generated by a vector field, ${\bf f}$:
\be
\label{map}
        \tilde{x}^\a = x^\a + \e f^\a(x^\b) \, ,
\ee
where $\e$ is an infinitesimal constant${}^e$.
The result of the mapping Eq. (\ref{map}) (see Fig.~3) is to move a
point $P(x^\a)$ to point ${P'}$ with coordinates $x^\a +
\e f^\a(x^\b)$.  Similarly, a neighboring point $Q(x^\a + d x^\a)$ will be
moved to point $Q'(x^\a + dx^\a + \e f(x^\a + dx^\a))$
(or, up to first order in differentials,
$Q'(x^\a + dx^\a + \e f^\a + \e{f^\a}_{,\g}dx^\g$).

The infinitesimal interval $\overline{P'Q'}$ is
\be
(x^\a + dx^\a + \e f^\a + \e {f^\a}_{,\g}dx^\g) - (x^\a + \e f^\a)
\ee
or
\begin{equation}
                     dx^\a + \e f^{\a}{}_{,\g}dx^\g \, .
\end{equation}
The length, $\overline{ds}^2$, of this interval is given by
\bea
\label{lie1}
\overline{ds}^2 &=&  g_{\a\b}(x^\g + \e f^\g)
                        (dx^\a + \e{f^\a}_{,\k} dx^\k)
                                (dx^\b + \e{f^\b}_{,\k} dx^\k)\nonumber\\
                &=&  (g_{\a\b} + \e g_{\a\b,\g} f^\g)
                                (dx^\a + \e{f^\a}_{,\g}dx^\g)
                                (dx^\b + \e{f^\b}_{,\k}dx^\k) \,.             
\eea
Expanding Eq. (\ref{lie1}), neglecting terms of order $\e^2$, and
rearranging dummy indices results in 
\be
\label{lie2}
\overline{ds}^2=g_{\a\b}dx^\a dx^\b + \e (g_{\a\b,\g}f^\g + g_{\a \g}f^\g_{,\b}
                              + g_{\g \b}{f^\g}_{,\a})dx^\a dx^\b\;.  
\ee
With the definition
\be
\label{lie3}
                2s_{\a\b} \equiv g_{\a\b,\g}f^\g + g_{\a\g}{f^\g}_{,\b}  + g_{\g\b}{f^\g}_{,\a}  
\ee 
the change in the metric may be written as
\be
          \frac{1}{\e}(\overline{ds}^2 - ds^2)= 2s_{\a\b}dx^\a\, dx^\b\;.
\ee
Because the left hand side of this equation is a scalar and $dx^\a\, dx^\b$ is a 
symmetric tensor it may be inferred that the symmetric quantity $s_{\a\b}$ is a 
covariant tensor${}^6$.  
 
The structure of the right hand side of Eq. (\ref{lie3})
\be 
               g_{\a\b,\g}f^\g + g_{\a\g}{f^\g}_{,\b}  + g_{\g\b}{f^\g}_{,\a}  
\ee
combines partial differentiation and a
vector field and arises in precisely this form in many applications.  The operation is important enough to have its own name and symbol.  It
is called the \lq\lq Lie derivative\rq\rq\ of a geometric object (in this instance, the metric tensor, $g_{ab}$) with respect to a vector field, $f$, 
and is written ${\cal{L}}_f$.  Thus, Eq. (\ref{lie3}) may be written as
\be
     {\cal{L}}_f g_{\a\b} = 2s_{\a\b}\;.    
\ee

In the special case where the metric tensor is invariant under the transformation we will have
${\cal{L}}_f g_{\a\b} = 0$.  Then the vector ${\bf f}$ is, by definition, a {\it Killing
vector}.  Killing vectors are customarily designated by the symbol \(\mbox{\boldmath $\x$}\).  For Killing vectors, then, we have 
\be
\label{lie4}
  {\cal{L}}_{\x} g_{\a\b} =  g_{\a\b,\g}\x^\g + g_{\a\g}{\x^\g}_{,\b}  + g_{\b\g}{\x^\g}_{,\a} = 0  \,.      
\ee
This equation can be written in a different form, useful for many purposes, {\it viz.},
\be
\label{kil}
          {\cal{L}}_{\x} g_{\a\b} = \x_{\a;\b} + \x_{\b;\a} = 0 \;.  
\ee 
The identity of these 2 forms can be demonstrated by expressing both in geodetic coordinates.  The result is
identical expressions.  Consequently, they are identical in any choice
of coordinates.  Alternatively, expansion of the covariant derivatives
of Eq. (\ref{kil}) yields precisely Eq. (\ref{lie4}).  Any 2-index tensor may be expressed as the sum of a symmetric and skew-symmetric tensor.
Equation (\ref{kil}) indicates that the symmetric part of the tensor ${\x_\a}_{;\b}$
vanishes.  Because infinitesimal displacements generated
by Killing vectors leave the metric unchanged, these displacements map geodesics
onto neighboring geodesics.  Note, also, that Eqs. (\ref{lie3}) may be read in either of two ways.
Given a metric, they provide the means for the determination of its
Killing vectors or given a set of Killing vectors, determining the
metric.  The latter is ill-defined. 
\section{Killing Vectors in Flat Spacetimes}
\setcounter{equation}{0}
In flat spacetimes Cartesian coordinates may be introduced.  Consequently, in Eqs. (\ref{kil}),
covariant derivatives may be replaced by partial derivatives and the right hand portion 
becomes
\be
\label{mink1}
         \x_{\a,\b} + \x_{\b,\a} = 0 \;.  
\ee                     
By differentiation we get 
\be
\label{mink2}
         \x_{\a,\b,\g} + \x_{\b,\a,\g} = 0   
\ee
and by cyclic permutation of the indices we obtain 
\be
\label{mink3}
         \x_{\b,\g,\a} + \x_{\g,\b,\a} = 0   
\ee
and
\be
\label{mink4}
         \x_{\g,\a,\b} + \x_{\a,\g,\b} = 0 \,.   
\ee
Addition of Eqs. (\ref{mink2}) and (\ref{mink3}), subtraction of  Eq. (\ref{mink4}), and recognition that
second partial derivatives commute, results in the differential equations
\be
\label{mink5}
                      2\x_{\g,\a,\b} = 0\,.   
\ee
The solutions of these equations are the general linear functions
\be
\label{mink6}
                 \x_\a = A_{\a\b}x^\b + B_\a
\ee
where $A_{\a\b}$ and $B_\a$ are constants.  If Eq. (\ref{mink6}) is substituted into 
Eq. (\ref{mink1}) we find immediately that $A_{\a\b}$ is skew-symmetric 
\be
                                      A_{\a\b} = -A_{\b\a}\,.
\ee
We see, thus, that an n-dimensional flat space has $n(n+1)/2$ independent Killing vectors.  For Minkowski spacetime this is just $10$.
This demonstration depends in an essential way on the {\it flatness} of the spacetime.  It is this fact which permits the substitution of partial for covariant differentiation 
and consequent ability to reorder the differentiation.
\section{Conformal and Homothetic Motions}
\setcounter{equation}{0}
In addition to the mappings described by Killing vectors, there are
other classes of transformations generated by vector fields which are
important in the present context.  These are the so called \lq\lq
homothetic\rq\rq\ and \lq\lq conformal\rq\rq\space motions.  In
these cases, respectively, the metric tensor is either multiplied by a constant or a
scalar function.  The generators are termed homothetic or conformal Killing vectors.  
In these cases the stress tensor is {\it proportional} to the metric.  Both cases are subsumed in
\be
\label{lie5}
\x_{\a;\b} + \x_{\b;\a} = 
               g_{\a\b,\g}\x^\g + g_{\a\g}{\x^\g}_{,\b} + g_{\g\b}{\x^\g}_{,\a} =
                            2\sigma g_{\a\b}\;.
\ee

It is simple to determine $\sigma$.  Contract Eq. (\ref{lie5}) with
$g^{\a\b}$; simple manipulation yields $\sigma = {\x^\g}_{;\g}/n$ and
\be
\label{lie6}
g_{\a\b,\g}\x^\g + g_{\a\g}{\x^\g}_{,\b} + g_{\g\b}{\x^\g}_{,\a} =
          \frac{2}{n}{\x^\g}_{;\g} g_{\a\b}
\ee
where $n$ is the dimension of the manifold.
If ${\x^\g}_{;\g}$ is a constant, the Killing vector,
by definition, describes homothetic motions;
and if ${\x^\g}_{;\g}$ is a scalar field, say $\f(x^\a)$,
the motion is called conformal.

Similarly to Eqs. (\ref{kil}) this may be written as
\bea
         \x_{\a;\b} + \x_{\b;\a} & = & \frac{2}{n}{\x^\g}_{;\g} g_{\a\b}
         \;\;\; {\rm or} \nonumber\\ [3mm]
\label{kil2}         {\cal{L}}_{\x} g_{\a\b} & = &  \frac{2}{n}{\x^\g}_{;\g} g_{\a\b}
\eea
In this instance the trace-free symmetric part of  \( \x_{\a;\b}\) vanishes.  

Again, for flat spacetimes, the situation is vastly simplified.  A special homothetic Killing vector
is given by
\be
\label{kilflat}
                        \x^\a = \k x^\a 
\ee
where $\k$ is a constant.  (See Fig. 4.)  The most general homothetic Killing vector is
\be
\label{kil2flat}
                  \x_\a = (\k\eta_{\a\b} + A_{\a\b})x^\b +B_\a 
\ee
where $A_{\a\b} = -A_{\b\a}$.
This is readily verified by substitution into in Eq. (\ref{lie5})
and replacing covariant by partial derivatives.  In this instance 
the coordinate grid is uniformly stretched or shrunk.
Similarly, it is easy to confirm that 
\be
                 \x_\a = (\eta_{\a\b} C_\g - \frac{1}{2}\eta_{\b\g} C_\a)x^\b x^\g
\ee
where $\eta_{\a\b}$ is the flat space metric and $C_\g$ are constants, are special conformal Killing vectors.
\section{Redshifts Derived from Killing Vectors}
\setcounter{equation}{0}
The calculation of redshifts is extremely simple in spacetimes possessed of time-like 
conformal Killing vector fields parallel to the world lines of source and observer.
Consider Eq. (\ref{weyl}), Weyl's universal definition of redshift,
\be
                  1 + z = \frac{ds'}{ds}\, .
\ee 
The event $P(x^\a)$ at the source is connected to the event $Q(y^\a)$ at the observer by a null
geodesic.  (See Fig~4.)  A conformal Killing vector field $\x^a$ moves $P(x^\a)$ into $P'(x^\a +\e\x^a(x^\b))$and
$Q(y^\a)$ into $Q'(y^\a +\e\x^a(y^\b))$ and the null geodesic connecting $P$ and $Q$ into the null geodesic
connecting $P'$ and $Q'$.  We have then respectively
\be
\label{red1}
                  ds = \mid \e \x^a(x^\b)\mid \;\;\;\;{\rm and}\;\;\;\; ds' = \mid \e \x^\a(y^\b)\mid
\ee
and thus
\be
\label{red2}
                  1+z = \sqrt{\frac{g_{\a\b}(y^\g)\x^\a(y^\g)\x^\b(y^\g)}
                             {g_{\a\b}(x^\g)\x^\a(x^\g)\x^\b(x^\g)}}            \,.
\ee
This holds if the Killing vector field is tangent to the world line of the source 
at its point of emission and tangent to the observer's world line at the point of reception.  
If source and observer move on conformal time-like Killing lines this condition is fulfilled 
at all times.
\section{Doppler Effect in Minkowski Spacetime}
\setcounter{equation}{0}
Weyl's definition of redshift, $1 + z = ds'/ds$, readily provides the usual formula for
relativistic Doppler effect in flat spacetime.  Take the world line of the source to be 
\be\begin{array}{ccccccc}
                            x^0 &=& t & &   &   &      \\
                             x  &=& y &=& z & = & 0\,.

\end{array}
\ee

For the world line of the (inertial) observer 
we take a straight line through the origin with slope $\b$  
\be
                x= \b t,\;\;\;y = z = 0 \,.
\ee
Light emitted by the source at time $t$ will be received by the observer at time $t'$.  (See Fig.~5.)  If
the observer is receding from the source he or she will see the light at 
\be
                x' = \b t'\;\;\;y' = z' = 0 \,.
\ee
The homothetic Killing vector (\ref{kilflat}) has the lengths, respectively, at source 
and observer of $\k t$ and $\k(t^{'2} - x^{'2})^{1/2}$.  We readily obtain from Eq. (\ref{red2})
\bea
\label{redshift}
                    1+z &=& \sqrt{\frac{g_{\a\b}(y^\g)\x^\a(y^\g)\x^\b(y^\g)}
                               {g_{\a\b}(x^\g)\x^\a(x^\g)\x^\b(x^\g)}} \nonumber \\
           &=&               \frac
                                       { \sqrt{{t'}^2 - {x'}^2}}
                                                    {t}
                                                                     \nonumber \\  
           &=&  \frac
                                   {t'\sqrt{1-\b^2}}
                                      {t' - x'}
                                                                    \nonumber  \\
           &=& \frac
                                   {\sqrt{1-\b^2}} 
                                      {1 - \b}
                                                                   \nonumber \\   
           &=& \sqrt{\frac{1+\b}{1-\b}}
\eea
This is the relativistic Doppler formula for a receding observer.  For an approaching observer 
replace $\b$ by $-\b$.

It is instructive to apply Weyl's definition of redshift, Eq. (\ref{weyl}) to Minkowski spacetime with the spatial part expressed in polar spherical coordinates.
\be
\label{polar}
       ds^2 = dt^2 - dr^2 -r^2(d\th^2 + \sin^2\th\, d\f^2)\,.
\ee
We introduce {\it 4-dimensional\/} polar coordinates with the coordinate transformation
\be
\label{sub}
                \left( \begin{array}{c} t\\r \end{array} \right) =     
                \left( \begin{array}{c} T \cosh\chi \\
                                   T \sinh\chi \end{array} \right) \,.
\ee
The differentials are
\bea
\label{diff}
                  dt = dT\cosh\chi + T\sinh\chi d\chi \nonumber \\
                  dr = dT\sinh\chi + T\cosh\chi d\chi 
\eea
The result is
\be
\label{milne}
                   ds^2 = dT^2 - T^2\,[d\chi^2 + \sinh^2\chi\ (d\th^2  + \sin^2\th\, d\f^2)]\,. 
\ee
This maps the interiors of the past and future light-cones of Minkowski spacetime into, respectively, linearly contracting and expanding spaces of constant negative curvature.  (See Fig.~6.)
A straight world line through the origin would be just
\be
                               \b = \frac{r}{t}
\ee
which by virtue of Eqs. (\ref{sub}) would map into 
\be
                               \b = \tanh\chi\,.
\ee
In the mapped space, such lines are, 
at constant $\{\chi,\,\th,\,\f\}$ with $\chi$ as the {\it rapidity}.
We now use the homothetic Killing vector Eq. (\ref{kilflat}) to construct a linear first-order differential form
\be
\label{minkform}
                         \x_\a  dx^\a = \k x_\a dx^\a = \k (t\,dt - r\,dr)\,.
\ee
In Eq. (\ref{milne}), with use of Eqs.(\ref{diff}), this is readily found to be
\be
\label{milform}
                        \x_\a  dx^\a = \k TdT\,.
\ee
The length of the homothetic killing vector in the new coordinate system is thus $\k T$ and the redshift is
\be
                          1 + z = \frac{T'}{T}\,.
\ee
%
\section{More Redshifts}
\subsection{Cosmological Redshifts}
\setcounter{equation}{0}
The last example leads us directly to the calculation of redshift in the Friedmann cosmological models${}^{7,8}$
\be
\label{fried}
                    ds^2 = dt^2 - R^2(t)[d\chi^2 + S^2(\chi)(d\th^2 + \sin^2\th d\f^2)]
\ee
where $S$ determines the curvature of the 3-space,
\be         S =      \left\{ \begin{array}{rl}
                                   \sin\chi \, ,& {\rm positive} \\
                                       \chi \, , &  {\rm flat} \\
                                  \sinh\chi \, , &  {\rm negative}
                       \end{array}        \right\} \,.
\ee
With Eq. (\ref{lie6})
\be
            g_{\a\b,\g}\x^\g + g_{\a\g}{\x^\g}_{,\b} + g_{\g\b}{\x^\g}_{,\a} =
          \frac{2}{n}{\x^\g}_{;\g} g_{\a\b}
\ee
it is readily confirmed that 
\be
                       \x^\a = R(t){\d^\a}_0
\ee
is a conformal Killing vector tangent to the world line of the source at $\chi = 0$ and also
tangent to the observer's world line at $\{\chi,\,\theta,\,\phi = constant\}$.

It follows from application of Eq. (\ref{red2}) that the redshift is 
\be
                         1 + z = \frac{R(t')}{R(t)}
\ee
for a light signal emitted at time $t$ and received at $t'$.  
\subsection{Redshifts in Stationary Spacetimes}
The general stationary spacetime may be written as
\be
\label{rw}
                 ds^2= g_{oo}dt^2 + 2g_{oi}dt dx^i + g_{ij}dx^i dx^j 
\ee
with $\partial{g}_{\a\b}/\partial{t}  =  0 $.

Because the field is stationary it necessarily possesses a time-like Killing vector.  The simplest
assumption is
\be
\label{rs9}
                               \x^\a = {\delta^\a}_0 \,.
\ee
This is readily verified by use of Eq. (\ref{lie4})
\be
                 g_{\a\b,\g}\x^\g + g_{\a\g}{\x^\g}_{,\b}  + g_{\b\g}{\x^\g}_{,\a} = 0 \,.
\ee

We then have, as earlier,
\bea
\label{stat}
                    1 + z &=& \sqrt{\frac{ g_{\a\b}(x'){\x^\a}(x')\x^\b{(x')} } 
                                         { g_{\g\d} (x){\x^\g}(x)\x^\d{(x)}}}\nonumber \\ 
                          &=& \sqrt{ 
                                     \frac{g_{oo}(x')} 
                                          {g_{oo}(x)} 
                                    }\,.
\eea
If the world line of the source is $x^j = const$ and that of the observer is ${x'}^j~=~const$, both are tangent to a Killing vector of this field.

One of the most important of the stationary metrics is the Schwarzschild-Droste (SD) metric
It is scarcely known among
relativists that the determination of the metric for a point mass was accomplished almost
simultaneously with Karl Schwarzschild by Johannes Droste, a Ph.D. 
student of H. A. Lorentz.  The form of the metric (\ref{dros}) was actually due to Droste${}^9$.  
\be
\label{dros}
   ds^2 = \left(1-\frac{2m}{r}\right)dt^2 - 
                       \left(1-\frac{2m}{r}\right)^{-1} dr^2
                                -r^2(d\th^2 + \sin^2{\th}d\f^2)\;.
\ee
By virtue of the Birkhoff theorem${}^{10,11}$, all spherically
symmetric, vacuum metrics are stationary and equivalent to Eq. (\ref{dros}).
Application of Eq. (\ref{stat}) yields the redshift
\be
                1 + z = \sqrt{\frac{1-2m/r'}{1-2m/r}} \,.
\ee
For the case where $r' = r + \delta r$ with $\delta r << r$ we easily obtain

\be
                        z= \frac{\delta\nu}{\nu} = -\frac{G\,M\,\delta r}{c^2\,r^2}\;.
\ee

In the usual special relativistic approximation${}^{12}$ the \lq\lq
mass\rq\rq\space of the photon is taken to be $h\nu /c^2$ and the
change in energy as it moves through a vertical distance $\delta r$ is
\be
                    \Delta E = \frac{h \nu}{c^2} \, g \, \delta r
\ee
where $g$ is the local gravitational acceleration ${GM}/{r^2}$.  
\subsection{Comments}
Both the preceeding subsections discuss \lq\lq elementary\rq\rq\ situations.  In the cosmological case the emitter is at rest in the coordinate system given by the metric (\ref{rw}).  In physical terms
terms it is at rest with respect to the background microwave radiation or equivalently, 
has no \lq\lq peculiar\rq\rq\ motion.  If the emitter does have a peculiar velocity the situation is substantially more complicated.  Moreover the usual observers are either earthbound astronomers or the Hubble Telescope which provides a peculiar motion at the receiving end.  But, this is quite small relative to other sources of error.

In the case of redshift due to differences in gravitational potential of emitter and observer in a stationary 
spacetime the situation is more clearly defined.  An example is the Pound-Rebka experiment${}^{13}$ where the difference is
precisely known and both emitter and observer are at rest.  A different and substantially more complex situation is
is presented by the Global Positioning System${}^{14}$.  Here, the observers are at rest and the emitters are not only at a
different gravitational potential, they are moving with high velocity
If a metric has Killing vectors in addition to the ones discussed for the radial Doppler effects they can be used for describing Doppler effect due to relative velocities of sources and observers with respect to a distinguished time-like
direction given by a (conformal) Killing vector defining a local state of rest.

\section{Conservation Theorems}
\setcounter{equation}{0}
One of the most important properties of Killing vectors is their
utility in the derivation of conservation theorems.  These are obtained in conjuction with
the tangent vectors of geodesics.  These are the null vectors for photons and the tangent vectors 
for force-free point masses.  The geodesics are provided
by the solutions of the geodesic equation
\begin{equation}
\label{geod}   
  \frac{dk^\a}{d\l} + \left\{
                                  \begin{array}{c}
                                     \a \\ \b \;\; \g
                                  \end{array}
                                \right\}       k^\b k^\g = 0
\end{equation}
where $\l$ is an {\it affine parameter} ${}^{15}$ along the trajectory and $k^\a = dx^\a /d\l$.
Note that Eq. (\ref{geod}) may be written as
\be
\label{geod2}
                             {k^\a}_{;\b}k^\b = 0\,.
\ee
Also, for photons and unit masses we have, respectively
\be
\label{geod3}                
                             {k^\a}k_\a = 0 \;\; {\rm and} \;\; 1 \,.
\ee
For a Killing vector, \(\mbox{\boldmath $\x$}\) and the tangent vector, 
\(\mbox{\boldmath $k$}\), to a geodesic,
the product ${\cal E} = \x_\a  k^\a$ is constant along
the geodesic.  This product is constant because the directional
derivative of $\cal E$ along the geodesic vanishes.
\bea
\label{rs7}
\dot{\cal E}& \equiv & (k^\a\x_\a)_{;\b}k^\b  \nonumber \\
        & = & {k^\a}_{;\b}k^\b\x_\a + \x_{\a;\b} k^\a k^\b = 0 \;. 
\eea
On the right hand side the first and second terms vanish by virtue, respectively, of
the geodesic equation, Eq. (\ref{geod2}) and the skew-symmetry of
$\x_{\a;\b}$. (See Eq. (\ref{kil}).)  Consequently, $\dot{\cal E} = 0$
and $\cal E$ is constant along the geodesic.

Equation (\ref{rs7}) is valid for either photons or
point masses.  For photons there is
an additional possibility.  If the metric admits either a {\it
homothetic\/} or {\it conformal\/} Killing vector a similar integral
exists.  The second term in Eq. (\ref{rs7}) will vanish
because $\x_{\a;\b}$ is symmetric and proportional to $g_{\a\b}$ and $k^\a$
is a null vector.  In either event the integral is identical in form
to that for Killing vector fields, that is,
${\cal E} = \x_\a \, k^\a = constant$.

Killing vectors are indispensable for the invariant formulation of conservation theorems
for fields and extended bodies.  Local conservation laws are expressed as the covariant 
divergence of a symmetric tensor 
\be
                  {T^{\a\b}}_{;\b} = 0\,, \;\;\;\;\;  T^{\a\b} = T^{\b\a}\,.
\ee 
Given a Killing vector $\x_\a$ define the quantity $S^\b \equiv \x_a T^{\a\b}$.  We then have
\be
\label{con1}
                         {S^\b}_{;\b} = \x_{\a;\b} T^{\a\b} +  \x_\a {T^{\a\b}}_{;\b} =0\,.
\ee
The first term vanishes because skew-symmetric and symmetric tensors are contracted; the second term vanishes
by virtue of the definition of one of its factors.  Now, the covariant divergence of a vector may be written 
as${}^{16}$
\be
                          {S^\b}_{;\b} =  \frac{1}{\sqrt{-g}}
                                                      (\sqrt{-g}S^\b)_{,\b} = 0\,.
\ee   
which is a true conservation law.

If, in addition to being symmetric, $T^{\a\b}$ has a vanishing trace, that is $T^{\a\b}g_{\a\b} = 0$, as is the 
important case of electromagnetic energy-momentum tensor, then conservation laws involving conformal Killing vectors may be obtained.  
In the first term of Eq. (\ref{con1}) substitute in accordance Eq. (\ref{lie5}). 
This results immediately in
\be
                         {S^\b}_{;\b} =\sigma g_{\a\b} T^{\a\b} +  \x_\a {T^{\a\b}}_{;\b} =0 \,.
\ee
On the right hand side both terms obviously vanish.
%
\newline
\newline
\vspace{3mm}
${}^{a)}$Professor Emeritus, Queens College, City University of New York \\
Electronic mail: ${}^{b}$harvey@scires.acf.nyu.edu \\
${}^{c}$elschucking@msn.com \\
${}^{d}$surow@attglobal.net 
\vspace{3mm}\\
${}^e$The following conventions are used: $c = 1$ 
with metric signature [$1,\,-1,\,-1,\,-1$]; coordinate indices range from [$0-3$] and in arbitrary 
coordinate systems are designated by lower case Greek letters, \it i.e., [$\a,\,\b,\,\g,\, \dots\,$];  spatial 
indices range from [$1-3$] and are designated by lower case Roman case letters [$i,\,\,k,\,j,\,\dots\,$]; 
in Cartesian coordinate systems we use $[t,\,x,\,y,\,z]$;
partial and covariant differentiation are indicated, respectively, 
by commas or semicolons, {\it viz.\/}, ${f^\a}_{,\b} = \frac{\partial{f^\a}}{\partial{x^\b}}$
and
${f^\a}_{;\b} =  \frac{ \partial{f^\a}}{\partial{ x^\b}}
                      + \left\{
                                  \begin{array}{c}
                                     \a \\ \b \;\; \g
                                  \end{array}
                      \right\} f^\g $.
\vspace{3mm}\\
${}^1$A. Einstein, \lq\lq \"{U}ber das Relativit\"{a}ts Prinzip und die aus demselben gezogenen Folgerungen\rq\rq\ [On the Relativity Principle and the Conclusions drawn from it],
{\it Jahrbuch der Radioaktivit\"{a}t und Elektronik\/}, {\bf 4}, pp. 411--462 (1907).
An English translation is available in vol.2, \lq\lq Collected Papers of Albert Einstein\rq\rq, translator A. Beck, consultant P. Havas, Princeton University Press, Princeton, NJ, (1989) pp. 252--311.
\vspace{3mm}\\
${}^2$A. Einstein, \lq\lq\ Die Grundlage der allgemeinen Relativit\"{a}tstheorie,
{\it Annalen der Physik\/}, {\bf 49}, pp. 769--822 (1916).
An English translation is contained in vol. 6 \lq\lq Collected Papers of Albert Einstein\rq\rq, translator A. Engel,  consultant E. L. Schucking, Princeton University Press, Princeton, NJ, (1997)
pp. 147--200.
\vspace{3mm}\\
${}^3$H. Weyl, {\it Raum-Zeit-Materie}, 5th edition,
Springer-Verlag, Berlin (1923). This was edited and annotated by
Juergen Ehlers and reprinted as the 7th edition, Springer-Verlag,
Berlin (1988).  The English translation is by E. L. Schucking.
\vspace{3mm}\\
${}^4$See, e.g., R. d'Inverno, {\it Introducing Einstein's Relativity\/},
Clarendon Press, Oxford (1992), Section 7.7.
\vspace{3mm}\\
${}^5$S. W. Carroll, {\it Spacetime and Geometry\/}, Addison Wesley, New York (2004).  See section 3.8.
\vspace{3mm}\\
${}^6$In 3-dimensional elasticity theory this tensor is known as 
the strain tensor due to an infinitesimal displacement $\e f^i(x^j)$.  See, e.g., F. Ziegler, 
{\it Mechanics of Solids and Fluids\/}, Springer-Verlag, New York (1991).  
See section 1.3, \lq\lq Kinematics of Deformable Bodies\rq\rq.
\vspace{3mm}\\
${}^7$W. Rindler, {\it Relativity - Special, General, Cosmological\/}, Oxford University Press, New York (2001). See sections 16.4 and 16.5.
\vspace{3mm}\\
${}^8$S. W. Carroll, {\it loc cit}, section 8.2.
\vspace{3mm}\\
${}^9$A note by T. Rothman discussing its genesis and content has appeared as a \lq\lq Golden Oldie\rq\rq\ in {\it General Relativity and Gravitation\/}, {\bf 34}, pp.1541--1543 (2002).
\vspace{3mm}\\
${}^{10}$R. d'Inverno, {\it loc cit\/}, section 14.6.
\vspace{3mm}\\
${}^{11}$S. W. Carroll, {\it loc cit\/}, section 5.2.
\vspace{3mm}\\
${}^{12}$W. Rindler, {\it loc cit\/}, pp. 4--26.
\vspace{3mm}\\ 
${}^{13}$R. V. Pound and G. A. Rebka, \lq\lq Apparent Weight of Photons\rq\rq, {\it Physical Review Letters}, 
{\bf 4}, pp.~337--341 (1960).
\vspace{3mm}\\
${}^{14}$N. Ashby, \lq\lq Relativity and the Global Positioning System\rq\rq, {\it Physics Today\/}, May 2002
pp.~41-47; \lq\lq Relativity in the Global Positioning System\rq\rq, {\it Living Reviews in Relativity\/},
http://www.livingreviews.org/lrr--2003--1, Max-Planck-Gesellschaft, Potsdam, Germany (2003).
\vspace{3mm}\\
${}^{15}$S. W. Carroll, {\it loc cit\/}, section 3.4
\vspace{3mm}\\
${}^{16}$S. W. Carroll, {\it loc cit\/}, see p.~101.
\begin{figure}
{\bf \caption{Weyl's Geometry of the Doppler Effect}}
The points $P$ and $P'$ on the worldline of a point-like light source
are the origin of surfaces of constant phase that form future null cones.
The worldline of the observer intersects these two light cones
in points Q and $Q'$.
The light signals $PQ$ and $P'Q'$ are given by null geodesics
in the ray approximation.
We draw the light cones as having \lq\lq straight\rq\rq\ sides,
but that is strictly a convention;
the intervening spacetime could be curved.

{\bf \caption{Doppler Geometry}
{\rm The notation is the same as in Figure 1.
If the light-emitting process in the source has the infinitesimal period
lasting from events $P$ to $P'$ it will be perceived by the observer
of having a period lasting from $Q$ to $Q'$.
The ratio of their proper times $ds'/ds$ is the redshift $1 + z$.}}

{\bf \caption{Killing Motion}
{\rm The vector field $\xi^{\alpha}$ moves two neighboring points $P$ and $Q$
by $ \epsilon \xi^{\alpha} $ into the points $ P' $ and $ Q' $.
This will in general change the distance $ds$ of the infinitesimal interval
$ \overline{PQ} $ into the distance $ \overline{ds} $ of $\overline{P'Q'} $.
If $ \overline{ds} = ds $ for all neighboring points $Q$,
then $\xi^{\alpha}$ is a Killing vector field.}}

{\bf \caption{A Conformal Vector Field for Source and Observer}}
The conformal Killing vector field moves the null geodesic connecting $P(x)$ with $Q(y)$ 
into another null geodesic connecting $P'$ with $Q'$.
If source and observer move on conformal Killing lines,
the redshift $1 + z$ is given by
the ratio of the length of the Killing vectors $|\xi(y)|/|\xi(x)|$.

{\bf \caption{Radial Doppler Shift in Minkowski Spacetime}}
Coordinates are chosen to assume source and observer are in a 4-dimensional plane
with the source as time axis.
A light signal emitted by the source at time $t$ is received by the observer
at distance $x$ and time $t'$.
With speed of light $= 1$, we have $ x = t' - t$.

{\bf \caption{4-dimensional Polar Coordinates for Spacetime}}
Straight time-like world lines through the origin lie at constant rapidity $\chi$.
The transformation $ t = T \cosh \chi $, $ r = \sinh \chi $ gives $ T^{2} = t^{2} - r^{2} $.
$ T $ measures proper time along the rays originating from the origin.
$ T $ is also the length of the homothetic Killing vector.

\end{figure}
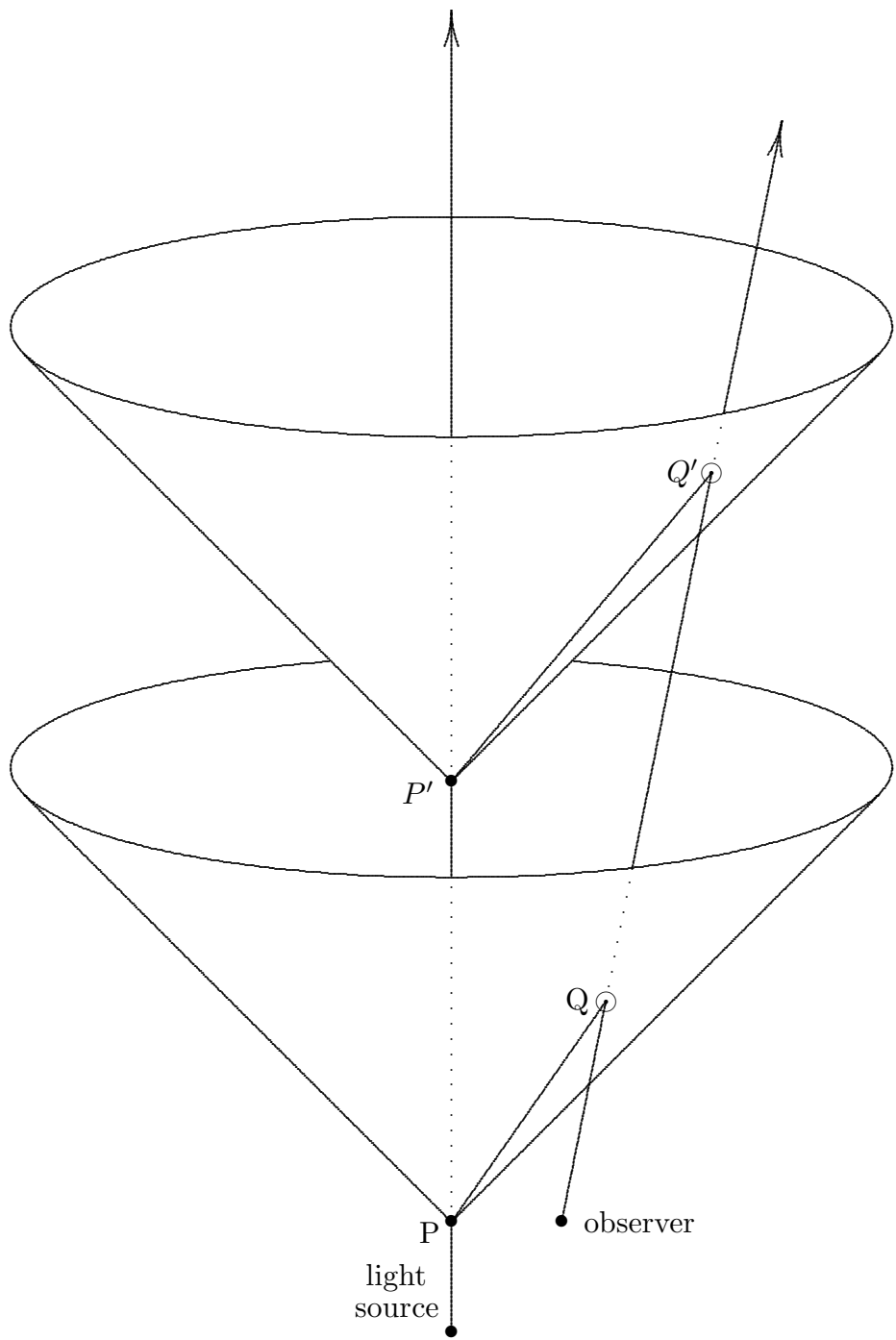
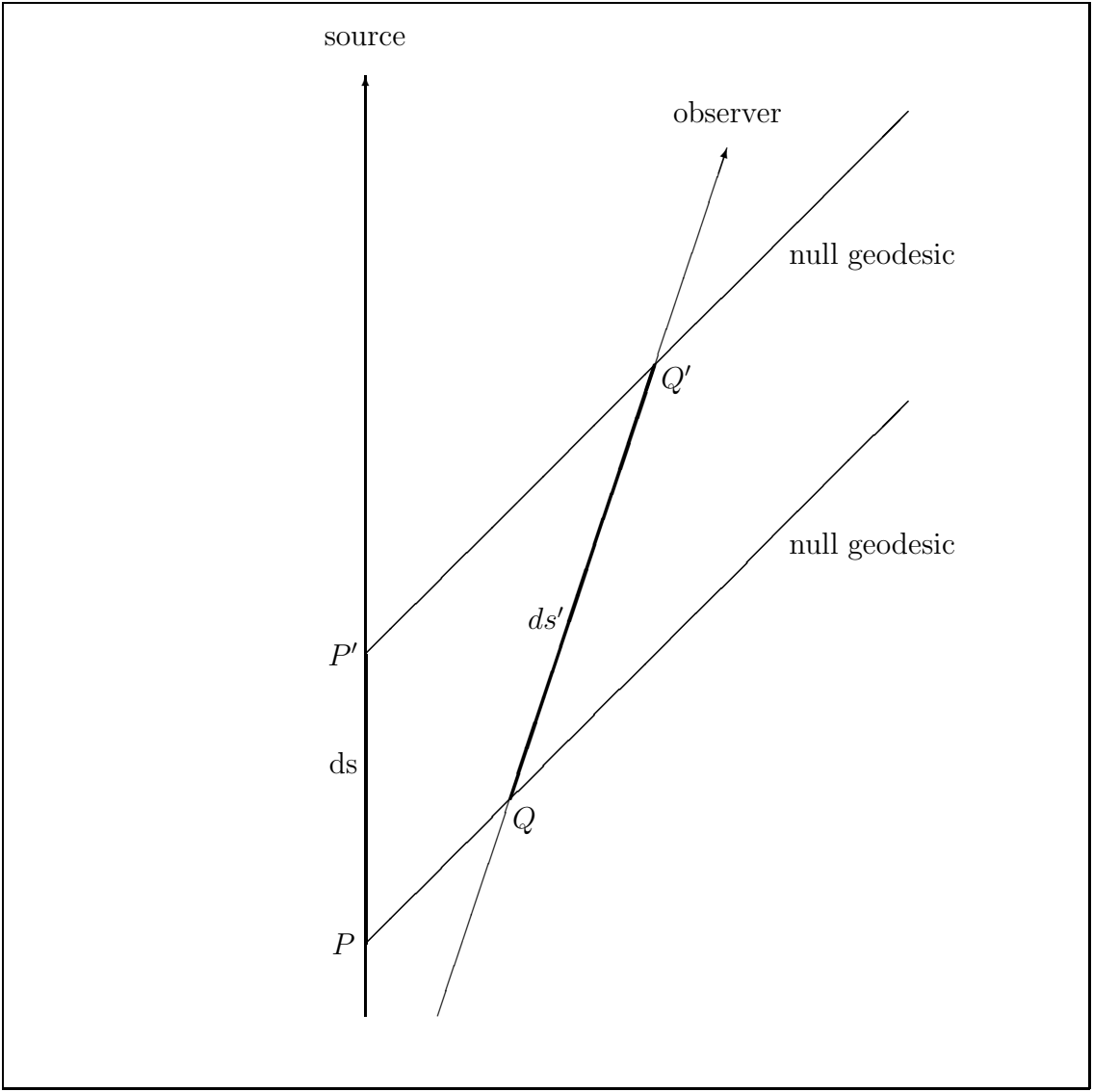
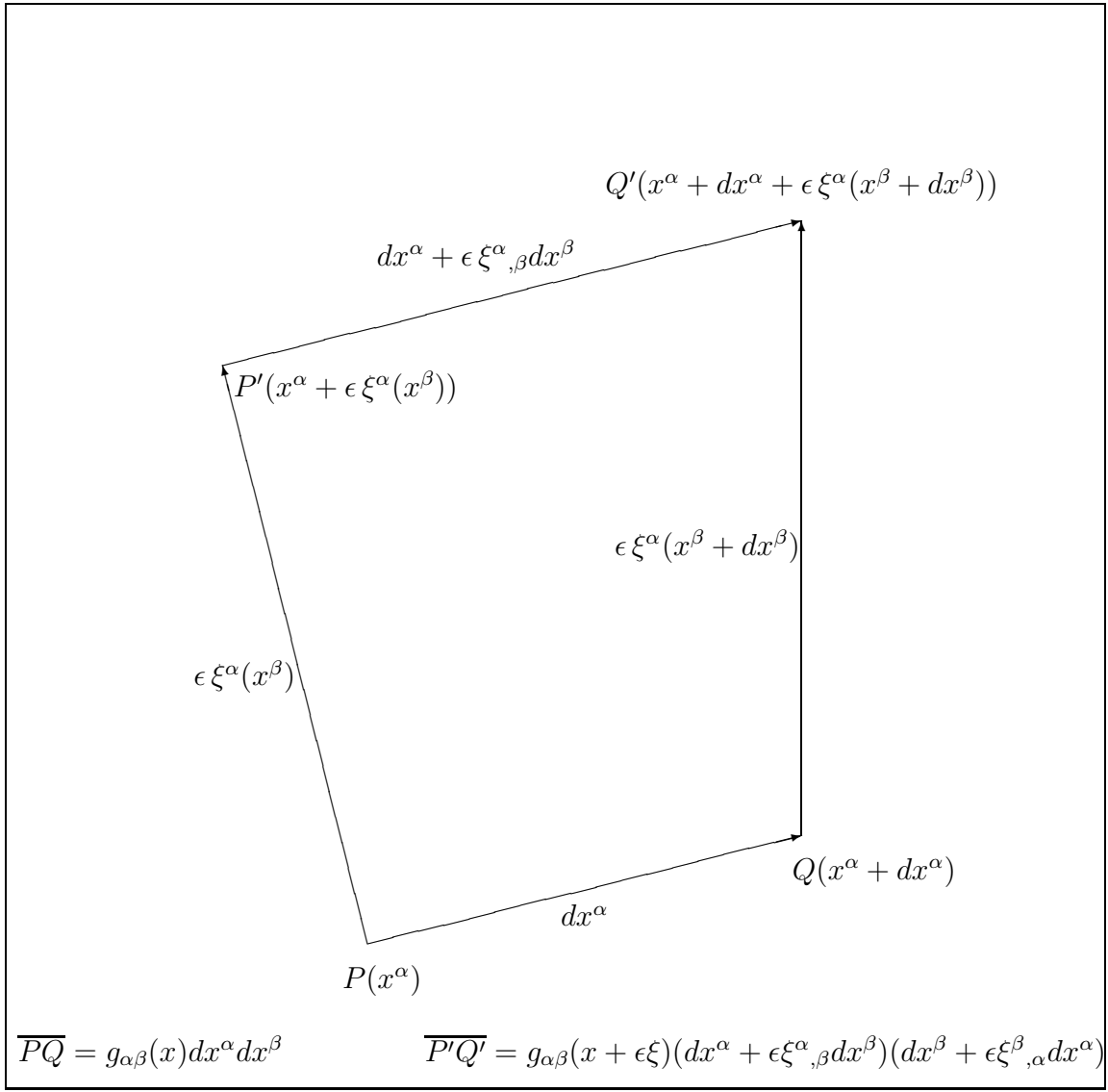
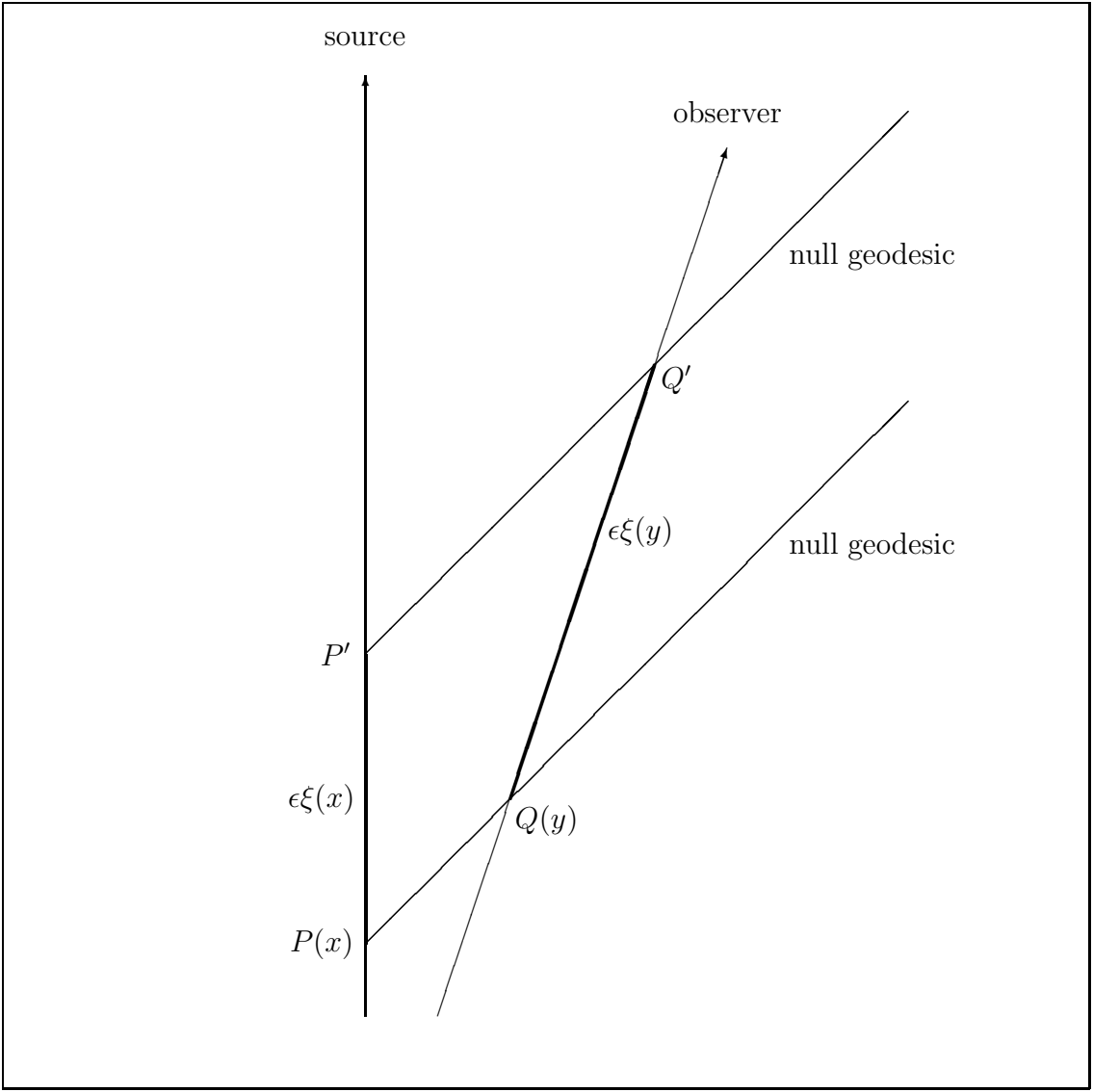
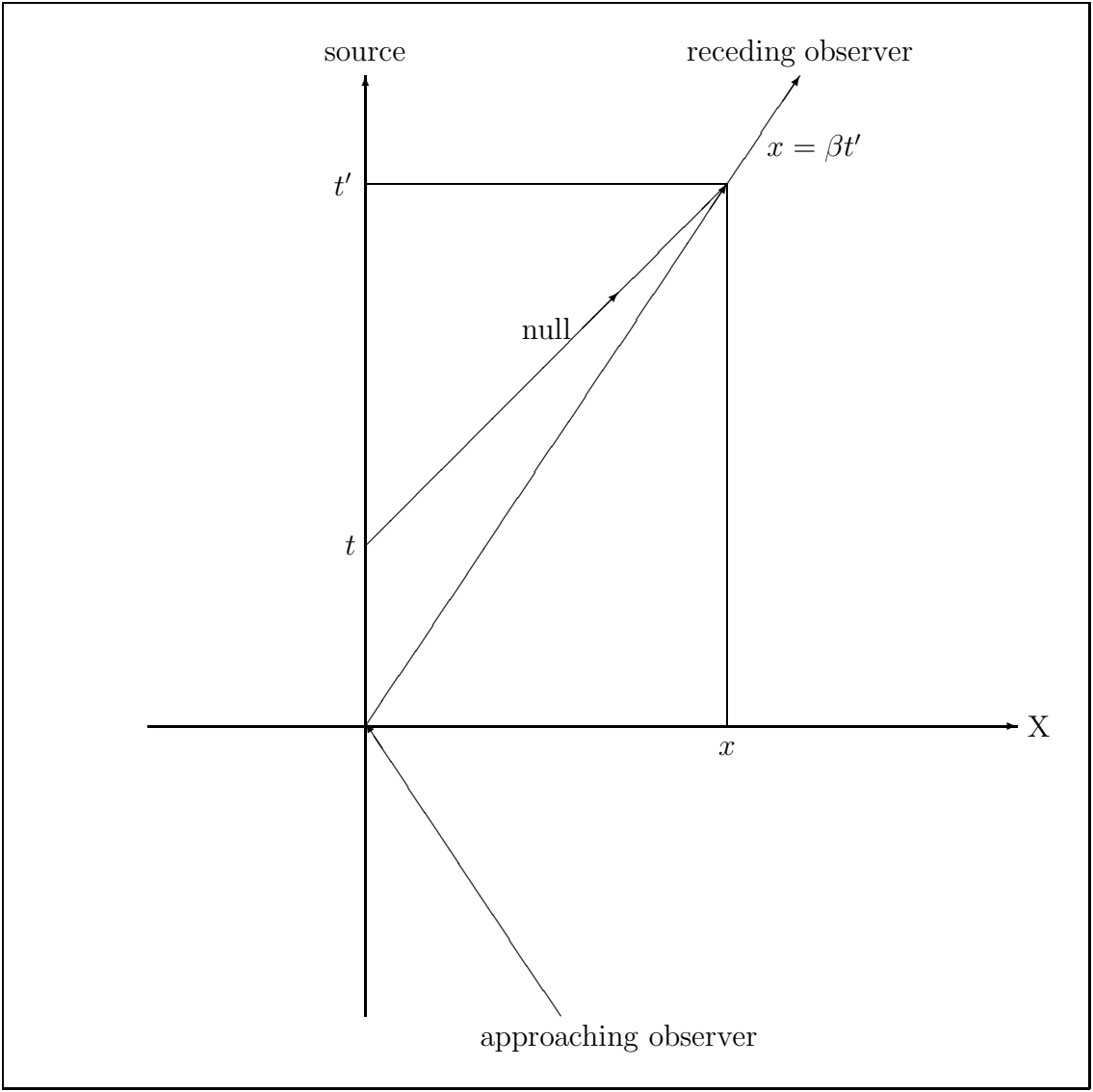
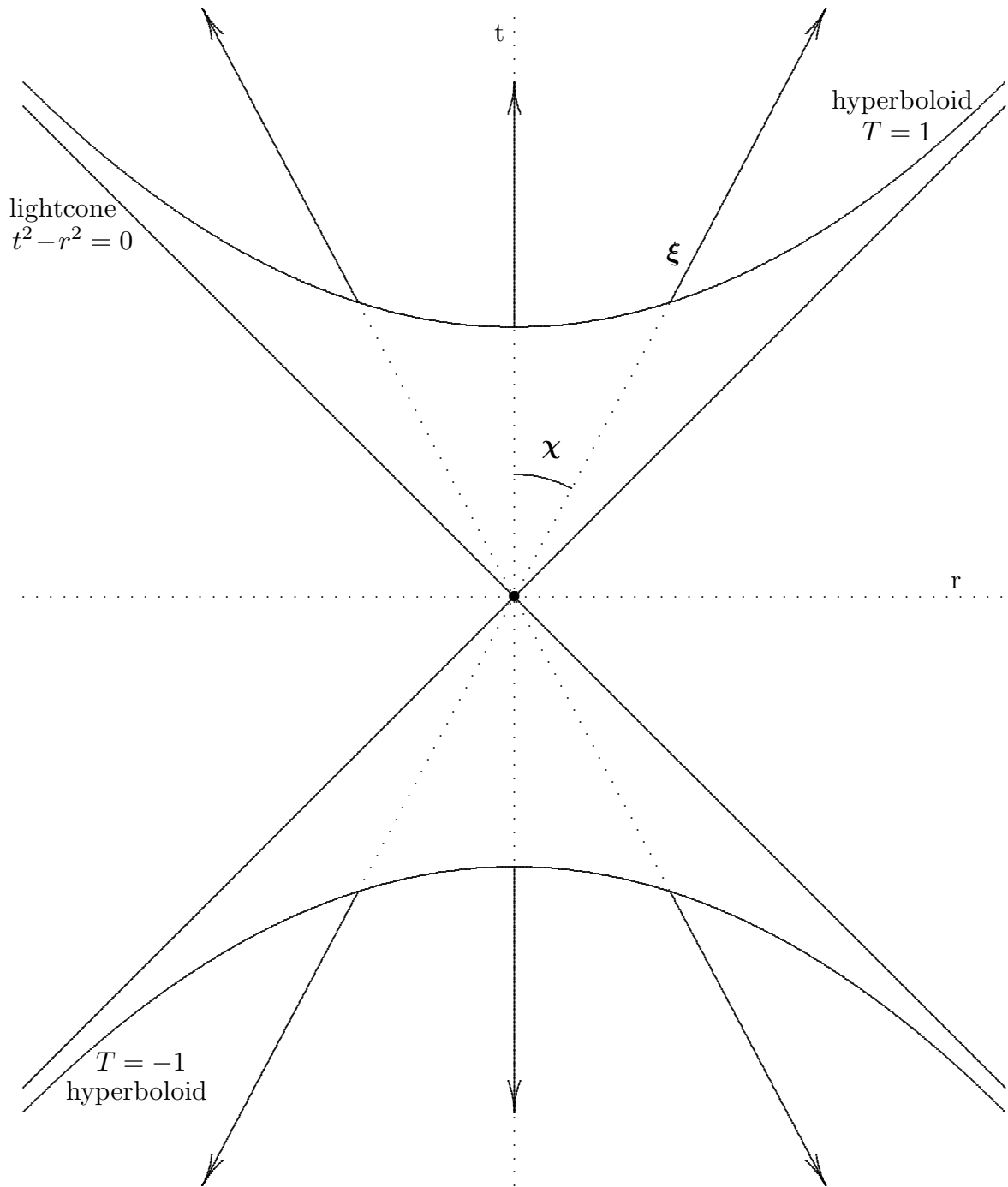
\end{document}